\documentclass[10pt,aps,prl,twocolumn,superscriptaddress,floatfix,longbibliography]{revtex4-1}

\usepackage{graphicx}
\usepackage{amsfonts}
\usepackage{amssymb}
\usepackage{amsmath}
\usepackage{txfonts}
\usepackage{lipsum}
\usepackage{color}
\usepackage{wasysym}
\usepackage[colorlinks=true, allcolors=blue]{hyperref}
\usepackage{bbold}
\usepackage{etoolbox} 
\usepackage[normalem]{ulem}
\usepackage{mathtools} 
\usepackage{multirow} 
\usepackage{hhline}
\usepackage{mathrsfs} 
\usepackage{soul}

\usepackage{letltxmacro}
\LetLtxMacro{\oldsqrt}{\sqrt}
\renewcommand{\sqrt}[2][\mkern8mu]{\mkern-6mu\mathop{}\oldsqrt[#1]{#2}}

\usepackage{url}
\definecolor{indigo(dye)}{rgb}{0.0, 0.25, 0.42}
\usepackage{placeins}

\begin{document}

\title{Can Orbital-Selective N\'eel Transitions Survive Strong Nonlocal Electronic Correlations?}

\author{Evgeny~A.~Stepanov}
\email{evgeny.stepanov@polytechnique.edu}
\affiliation{CPHT, CNRS, {\'E}cole polytechnique, Institut Polytechnique de Paris, 91120 Palaiseau, France}
\affiliation{Coll\`ege de France, Universit\'e PSL, 11 place Marcelin Berthelot, 75005 Paris, France}

\author{Silke~Biermann}
\affiliation{CPHT, CNRS, {\'E}cole polytechnique, Institut Polytechnique de Paris, 91120 Palaiseau, France}
\affiliation{Coll\`ege de France, Universit\'e PSL, 11 place Marcelin Berthelot, 75005 Paris, France}
\affiliation{European Theoretical Spectroscopy Facility, 91128 Palaiseau, France}

\begin{abstract}
Spin- or orbital-selective behaviours in correlated electron materials offer rich promise for spintronics or orbitronics phenomena and applications deriving from them.
Strong local electronic Coulomb correlations might lead to an orbital-selective Mott state, characterised by the coexistence of localized electrons in some orbitals with itinerant electrons in others.
Nonlocal electronic fluctuations are much more entangled in orbital space than the local ones. 
For this reason, finding orbital-selective phenomena related to nonlocal correlations, such as orbital-selective magnetic transitions, is a challenge.
In this work we investigate possibilities to realize an orbital-selective N\'eel transition (OSNT).
We illustrate that stabilising this state requires a decoupling of magnetic fluctuations in different orbitals, which can only be realized
in the absence of Hund's exchange coupling.
On the basis of two-orbital calculations for a Hubbard model with different bandwidths we show that the proposed OSNT can be found all the way from the weak to the strong coupling regime.
In the weak coupling regime the transition is governed by a Slater mechanism and thus occurs first for the narrow orbital.
At strong coupling a Heisenberg mechanism of the OSNT sets in, and the transition occurs first for the wide orbital.
Remarkably, at intermediate values of the interaction we find a non-trivial regime of the OSNT, where the Slater mechanism leads to a N\'eel transition occurring first for the wide orbital.
Our work suggests strategies for searching for orbital-selective N\'eel ordering in real materials, in view of possible spin-orbitronics applications.
\end{abstract}

\maketitle

The most striking effect of electronic Coulomb correlations in strongly correlated materials are probably phase transitions to various ordered states induced by collective electronic behaviour.
Strong local Coulomb repulsions between electrons favor localisation of the electrons on atomic sites and can drive the system toward a Mott-insulating state~\cite{Mott, RevModPhys.70.1039}.
Nonlocal collective electronic fluctuations are responsible for other types of orderings, in particular magnetic or superconducting states. 
In multi-orbital systems, the orbital degrees of freedom do not only have the potential to enhance these effects, but can also enable the emergence of non-trivial states of matter that cannot be realized in the single-orbital case.
Celebrated examples are orbital-selective states characterized by the coexistence of radically different collective electronic behaviors associated to distinct orbitals.
Such phenomena have attracted tremendous attention since the experimental observation of non-Fermi liquid behavior in the resistivity and an enhanced spin susceptibility in the metallic phase of the doped (${0.2<x<0.5}$) ruthenate Ca$_{2-x}$Sr$_{x}$RuO$_4$~\cite{PhysRevLett.84.2666}.
The material was suggested to undergo an orbital-selective metal-insulator transition (\mbox{OSMIT}) to a phase, where itinerant electrons in some orbitals coexist with localized electrons living in other orbitals~\cite{anisimov2002orbital}.
Such “orbital-selective Mott transitions” have instantaneously become a hot topic of condensed matter physics, triggering enormous excitement not only for ruthenate compounds~\cite{PhysRevLett.90.137202, PhysRevLett.93.177007, PhysRevLett.95.196407, PhysRevLett.96.057401, PhysRevLett.97.106401, PhysRevLett.102.086401, PhysRevLett.103.097001}, but also for iron-based superconductors~\cite{PhysRevLett.110.067003, 10.1038/nphys2438, 10.1038/s41535-017-0059-y, 10.3389/fphy.2021.578347}, iron chalcogenides~\cite{yi2015observation, huang2022correlation, 2023arXiv230405002K}, and the van der Waals ferromagnet Fe$_{3-x}$GeTe$_2$~\cite{PhysRevB.106.L180409}.
The link between theoretical findings in model systems~\cite{anisimov2002orbital, PhysRevB.69.045116, PhysRevLett.92.216402, PhysRevB.72.045128, PhysRevB.72.081103, PhysRevB.72.201102, PhysRevB.72.205124, PhysRevB.72.205126, PhysRevLett.95.116402, PhysRevLett.95.206401, PhysRevB.73.155106, PhysRevLett.99.126405, PhysRevLett.99.236404, PhysRevB.79.115119, PhysRevLett.102.126401, PhysRevLett.102.226402, PhysRevB.80.115109, refId0, PhysRevLett.104.026402, PhysRevLett.107.167001, PhysRevB.83.205112, PhysRevB.84.195130, PhysRevB.86.035150, PhysRevB.86.174508, PhysRevLett.110.146402, PhysRevLett.112.177001, Wang_2016, PhysRevB.100.115159, PhysRevLett.101.126401, Hackl_2009, Kou_2009, PhysRevLett.105.107004, PhysRevB.81.245130, PhysRevB.82.045125, PhysRevB.84.054527, PhysRevB.81.220506, PhysRevB.84.020401, PhysRevB.85.035123, doi:10.1142/S0217984913300159, PhysRevB.79.014505}
and observations in real materials remains however controversial, even more as the orbital-selective Mott phase is a rather fragile state that is unstable against both local~\cite{PhysRevLett.95.066402} and nonlocal~\cite{PhysRevLett.129.096403} interorbital hopping processes, and can also be destroyed by strong magnetic fluctuations~\cite{PhysRevLett.129.096404}.

Due to technical limitations, the overwhelming majority of theoretical studies so far have focused on 
paramagnetic Mott states, driven by purely local electronic correlations.
Likely more relevant to real materials questions at low temperatures are
an entirely different kind of orbital-selective phases, which 
are states originating from orbital-selective magnetic fluctuations, and in the extreme case magnetic orderings. Speculations about the existence of such states have been spurred on by the vast literature on iron-based
superconducting materials~\cite{PhysRevLett.104.057002, PhysRevB.85.180405, yin2011kinetic, PhysRevLett.116.247001, benfatto2018nematic, herbrych2018spin, patel2019fingerprints, PhysRevB.102.054430, PhysRevB.104.125122, PhysRevLett.127.077204, PhysRevB.105.075119}, but to date no strategy for realizing even the simplest orbital-selective N\'eel
phase has been established.
Doing so
requires using more advanced theoretical methods that go beyond local approximations.
Taking into account spatial collective electronic fluctuations in a multi-orbital framework is computationally demanding, which greatly limits possibilities of
studying \mbox{OSMIT}s to a symmetry-broken magnetic state~\cite{PhysRevLett.101.126401, Hackl_2009, Kou_2009, PhysRevLett.105.107004, PhysRevB.81.245130, PhysRevB.82.045125, PhysRevB.84.054527, PhysRevB.81.220506, PhysRevB.84.020401, PhysRevB.85.035123, doi:10.1142/S0217984913300159, PhysRevB.79.014505}.
Nonlocal collective electronic fluctuations that would drive the magnetic OSMIT are strongly entangled in orbital space~\cite{PhysRevLett.127.207205, PhysRevLett.129.096404}, which additionally complicates realizing the orbital-selective magnetic state.
Existing theoretical studies of magnetic \mbox{OSMITs} mostly start from models that assume the existence of a local magnetic moment which is then explicitly introduced in the theoretical description~\cite{PhysRevLett.101.126401, Hackl_2009, Kou_2009, PhysRevLett.105.107004, PhysRevB.81.245130, PhysRevB.82.045125, PhysRevB.84.054527}.
While giving interesting insights into the consequences of orbital-selective magnetic moments, such approaches do not allow one to decide on their existence. 
On the contrary, it has been shown recently that the ordered magnetic state is formed simultaneously within all orbitals that are coupled by the local interorbital exchange interaction (Hund's rule coupling)~\cite{PhysRevLett.129.096404}.

There are only few works, where the \mbox{OSMIT} to the ordered magnetic state was addressed more accurately based on interacting electronic models~\cite{PhysRevB.81.220506, PhysRevB.84.020401, PhysRevB.85.035123, doi:10.1142/S0217984913300159}.
In this set of works the authors performed dynamical mean-field theory (DMFT)~\cite{RevModPhys.68.13} or mean-field calculations for a two-orbital Hubbard model on a square lattice.
Various ordered magnetic states were investigated either by introducing two sublattices or within the dynamical cluster approximation~\cite{RevModPhys.77.1027}.
The authors argue that having distinct band dispersions for different orbitals is crucial for realising the \mbox{OSMIT} to the ordered magnetic state, and the proposed mechanism is not sensitive to the strength of the Hund's coupling $J$~\cite{PhysRevB.85.035123}.
Interestingly, this conclusion seems to be in contradiction with the absence of an OSNT found recently in a similar system
in the case of a finite Hund’s coupling $J$~\cite{PhysRevLett.129.096404}. 

In this work using an advanced many-body approach that includes spatial fluctuations beyond DMFT we propose a mechanism for the orbital-selective N\'eel transition (\mbox{OSNT}) that can be realised in a system with different bandwidths in the absence of Hund's coupling for an arbitrarily strong local Coulomb interaction.
We find that the OSNT occurs differently in the weak- and strong-coupling regimes, which can be associated respectively with the Slater and Heisenberg mechanisms of the N\'eel transition.
Interestingly, despite the absence of Hund's coupling, the electrons in different orbitals still do interact by means of the interorbital local Coulomb potential. 
This interaction manifests itself in the Kondo screening of the local magnetic moment of the narrow orbital by itinerant electrons of the wide orbital, which results in a simultaneous formation of the local moment in both orbitals.

{\it Model and method.}
We consider a half-filled two-orbital Hubbard model on a cubic lattice described by the Hamiltonian:
\begin{align*}
H = \sum_{jj',l,\sigma} t^{l}_{jj'} c^{\dagger}_{jl\sigma} c^{\phantom{\dagger}}_{j'l\sigma} + \frac12 \hspace{-0.05cm} \sum_{j,\{l\},\sigma\sigma'} U_{l_1l_2l_3l_4} c^{\dagger}_{j l_1 \sigma} c^{\dagger}_{j l_2 \sigma'} c^{\phantom{\dagger}}_{j l_4 \sigma'} c^{\phantom{\dagger}}_{j l_3 \sigma}\,,
\end{align*}
where $c^{(\dagger)}_{jl\sigma}$ is the annihilation (creation) operator for an electron on the lattice site $j$, in orbital $l\in\{1, 2\}$, with spin projection $\sigma\in\{\uparrow, \downarrow \}$.
$t^{l}_{jj'}$ is the intraorbital ($l$) hopping amplitude between sites $j$ and $j'$.
We restrict ourselves to nearest-neighbor hoppings only, and choose the half-bandwidth of the narrow
band as our unit of energy, i.e. $t^1_{jj^{\prime}}=1/6$. The second band is double as wide with 
$t^2_{jj^{\prime}}=1/3$.
The interaction is parametrized in the Kanamori form that includes the intraorbital ${U_{llll}=U}$, interorbital ${U_{ll'll'}=U-2J}$, spin flip ${U_{ll'l'l}=J}$, and pair hopping ${U_{lll'l'}=J}$ terms.
$J$ is the Hund's rule coupling.

An accurate description of the N\'eel transition requires accounting for long-range magnetic fluctuations and their influence on the electronic excitations (see, e.g., Refs.~\onlinecite{PhysRevX.11.011058, PhysRevLett.129.107202, PhD_Vandelli} and references therein). 
In the multi-orbital framework both of these aspects can be consistently taken into account by an advanced many-body approach dubbed ``dual triply irreducible local expansion'' (\mbox{D-TRILEX})~\cite{PhysRevB.100.205115, PhysRevB.103.245123, 10.21468/SciPostPhys.13.2.036}, which extends
the ``TRILEX'' approach of Refs.~\onlinecite{PhysRevB.92.115109, PhysRevB.93.235124} to the dual fermion~\cite{PhysRevB.77.033101, PhysRevB.79.045133, PhysRevLett.102.206401, BRENER2020168310} and boson~\cite{Rubtsov20121320, PhysRevB.90.235135, PhysRevB.93.045107, PhysRevB.94.205110, PhysRevLett.121.037204, PhysRevB.99.115124, PhysRevB.100.165128, PhysRevB.102.195109, PhysRevB.105.155151} variables.
In this method, nonlocal collective electronic fluctuations are treated self-consistently~\cite{stepanov2021coexisting, PhysRevLett.129.096404, PhysRevLett.127.207205, PhysRevResearch.5.L022016, Vandelli, Maria} by performing a diagrammatic expansion around DMFT~\cite{RevModPhys.90.025003, Lyakhova_review}.

\begin{figure}[t!]
\includegraphics[width=1\linewidth]{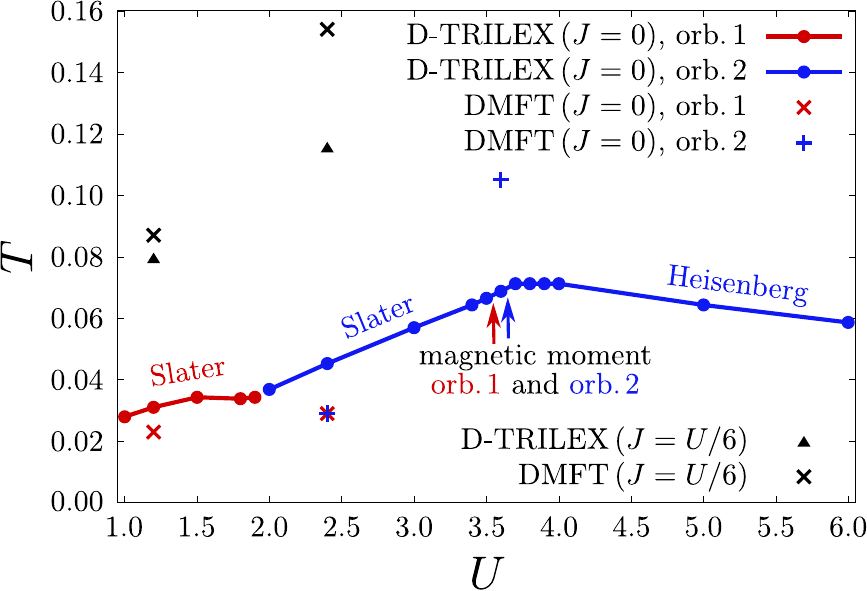}
\caption{N\'eel temperature for the two-orbital half-filled Hubbard model on a cubic lattice with different bandwidths of the two orbitals. Results are obtained using \mbox{D-TRILEX} and DMFT for different values of the Hund's exchange coupling $J$. At finite $J$ the N\'eel transition occurs simultaneously for both orbitals (black ``$\blacktriangle$'' and ``$\boldsymbol{\times}$'' markers). The OSNT scenario is realized in the absence of $J$: at ${U<1.95}$ upon decreasing the temperature the N\'eel transition occurs first for the narrow orbital (orb.\,1, red color), and at ${U>1.95}$ for the wide orbital (orb.\,2, blue color). Critical interactions at which the local magnetic moment is formed above the N\'eel transition are indicated by arrows.     
\label{fig:D-TRILEX}}
\end{figure}

{\it Results.} 
We have first performed \mbox{D-TRILEX} calculations for the two-orbital Hubbard model in the absence of Hund's coupling (${J=0}$). 
We calculate the orbital-selective N\'eel temperatures
by identifying the divergences of the orbital components of the static (${\omega=0}$) spin susceptibility ${X^{sp}_{ll'}({\bf q},\omega) = \langle m^{z}_{{\bf q},\omega,l} m^{z}_{-{\bf q},-\omega,l'} \rangle}$ obtained at the AFM wave vector ${{\bf Q}=\{\pi,\pi,\pi\}}$
\cite{PhysRevLett.129.096404}.
Here, ${m^{z}_{{\bf q},\omega,l}=\sum_{{\bf k},\nu,\sigma}c^{\dagger}_{{\bf k+q},\nu+\omega,l,\sigma}\sigma^{z}_{\sigma\sigma}c^{\phantom{\dagger}}_{{\bf k},\nu,l,\sigma}}$, and $\sigma^{z}$ is the familiar Pauli matrix.
The results are summarised in Figure~\ref{fig:D-TRILEX}. 

In the weak-coupling regime (${U<1.95}$) upon lowering the temperature the ${l=1}$ (narrow orbital) component of the AFM susceptibility diverges first, while the ${l=2}$ (wide orbital) component remains finite at the transition point.
This behaviour indicates the OSNT to a phase, where electrons in the narrow orbital 
order antiferromagnetically, while the wide orbital stays itinerant.
At ${U>1.95}$ we observe the opposite situation: the transition to the ordered AFM state occurs first for the wide orbital, while the narrow orbital remains itinerant.
Remarkably, the system exhibits an OSNT for any value of the interaction, except for ${U\simeq1.95}$ where ${T_{N_1}=T_{N_2}}$.

We now argue that the choice of vanishing Hund's coupling made above is essential for these results. Indeed,
in a multi-orbital system magnetic fluctuations of different orbitals are coupled due to the presence of Hund's exchange coupling $J$.
This coupling is realized through inter-orbital three-point (Hedin~\cite{GW1}) vertex corrections that are present in the self-energy and the polarization operator for any finite value of $J$.
These vertices connect the renormalized interaction in the spin channel to the electronic Green's function and thus are responsible for mixing different orbital contributions to the spin susceptibility.
Strong spatial magnetic fluctuations enhance this mixing, which leads to a simultaneous N\'eel transition for different orbitals~\cite{PhysRevLett.129.096404}.
Therefore, realizing the OSNT necessarily requires 
magnetic fluctuations of different orbitals to decouple.
This happens in the absence of Hund's coupling, since in this case the inter-orbital components of the vertex function in the spin channel are identically zero.
In realistic materials, the Hund's coupling can be suppressed, e.g., through the Jahn-Teller effect of phonons, which, as has been demonstrated for fullerides, can result even in a sign change of $J$~\cite{PhysRevB.62.7619, PhysRevLett.86.5361, doi:10.1126/sciadv.1500568, RevModPhys.81.943, PhysRevB.94.155152}.

\begin{figure}[t!]
\includegraphics[width=0.921\linewidth]{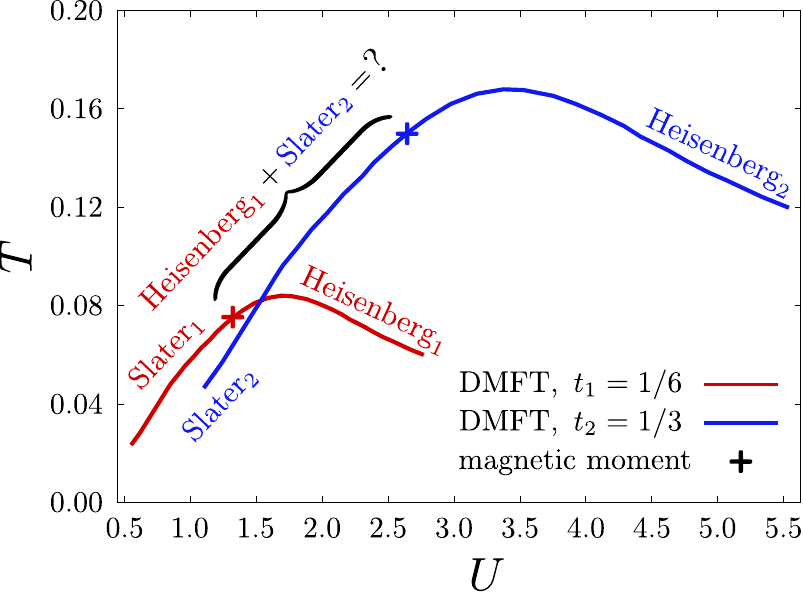}
\caption{Sketch of the proposed mechanism for the OSNT. Red and blue curves show the N\'eel phase boundaries predicted by DMFT for a half-filled single-band Hubbard model on a cubic lattice as a function of the local interaction strength $U$. Results are taken from Ref.~\onlinecite{PhysRevB.92.144409} and rescaled to get the phase boundaries for two different bandwidths defined by the nearest-neighbor hopping amplitudes ${t_1=1/6}$ (red) and ${t_2=1/3}$ (blue). ``$\boldsymbol{+}$'' markers indicate the point on each phase boundary at which the local magnetic moment is formed (see Ref.~\onlinecite{PhysRevB.105.155151}), which separates the Slater (at smaller $U$) and Heisenberg (at larger $U$) regimes of the N\'eel transition.     
\label{fig:DMFT}}
\end{figure}

Since at ${J=0}$ magnetic fluctuations of different orbitals decouple, the proposed mechanism of the OSNT can be qualitatively illustrated on the basis of single-orbital calculations.
Let us consider two separate single-orbital Hubbard models on a cubic lattice with different bandwidths defined by the nearest-neighbor hoppings ${t_1=1/6}$ (narrow orbital) and ${t_2=1/3}$ (wide orbital).
In Fig.~\ref{fig:DMFT} we compare the N\'eel temperatures $T_{N}$ for these two models as a function of temperature $T$ and interaction $U$. 
The N\'eel phase boundaries for the narrow (red) and wide (blue) orbitals are obtained by rescaling the results of DMFT calculations taken from Ref.~\onlinecite{PhysRevB.92.144409}.

Fig.~\ref{fig:DMFT} demonstrates that in the weak coupling regime the N\'eel temperature of the narrow orbital is larger than the N\'eel temperature of the wide orbital (${T_{N_1}>T_{N_2}}$).
However, in the strong coupling regime the relation between the N\'eel temperatures is opposite, namely ${T_{N_1}<T_{N_2}}$.
This result can be explained by the fact that in the Slater regime of magnetic fluctuation ${T_{N}}$ increases with increasing interaction. 
For a given value of $U$, the ratio $U$ over the bandwidth,
$U/W$, is stronger for the narrow orbital, which results in a higher $T_N$ for this orbital at weak-coupling.
In contrast, in the Heisenberg (strong-coupling) regime the N\'eel temperature is determined by the exchange interaction ${T_{N}\sim{}t^2/U}$, and the latter is larger for the wide orbital.

These two regimes of magnetic fluctuations can be distinguished by the absence (Slater) or presence (Heisenberg) of a local magnetic moment in the system.
In the single-orbital Hubbard model the formation of the local magnetic moment has been studied in Ref.~\onlinecite{PhysRevB.105.155151}. 
The critical point at the N\'eel phase boundary, where the local magnetic moment starts to form prior to the transition, is depicted in Fig.~\ref{fig:DMFT} by ``$\boldsymbol{+}$'' markers. It occurs close to the top of each dome-shaped curve.
Remarkably, at intermediate couplings (${1.5\lesssim{}U\lesssim1.7}$) one may expect a non-trivial regime, where the N\'eel transition occurs first for the wide orbital, where the local magnetic moment is not formed yet, while the narrow orbital remains itinerant but exhibiting a local moment.

\begin{figure}[t!]
\includegraphics[width=1\linewidth]{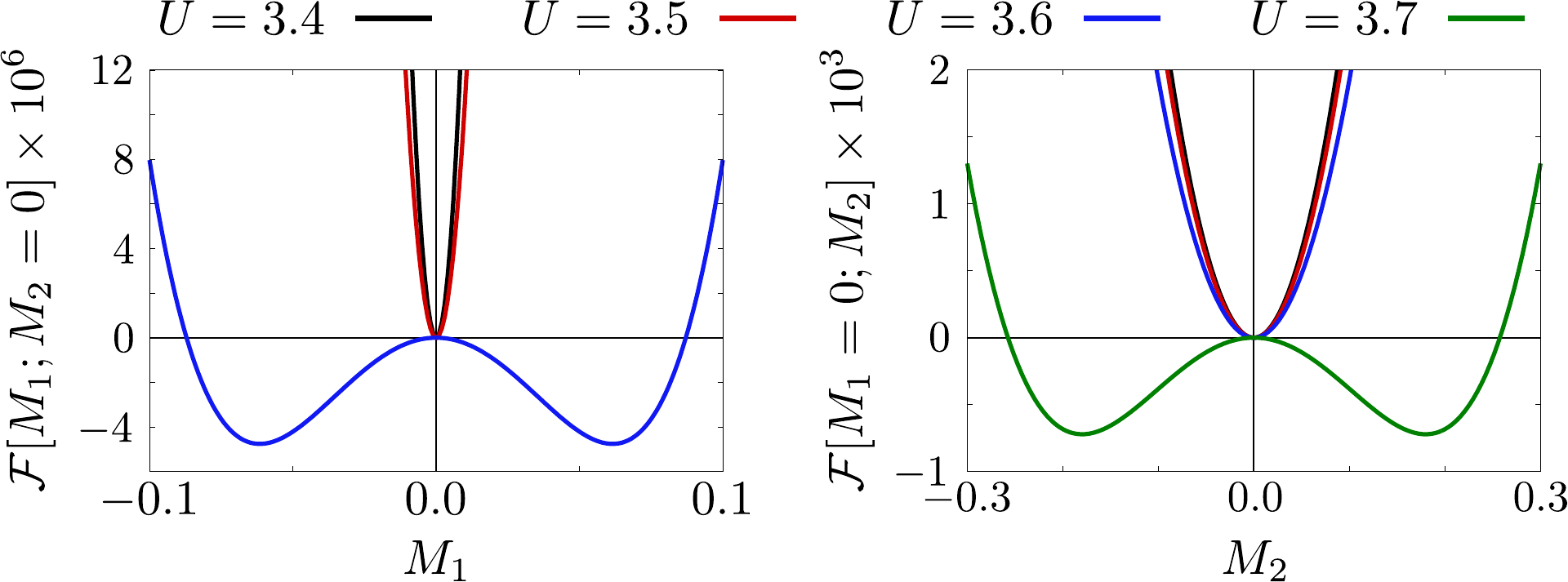}
\caption{Local free energy ${\cal F}[M_1;M_2]$ as a function of the value of one of the two local magnetic moments $M_1$ (orb.\,1, left panel) and $M_2$ (orb.\,2, left panel) obtained along the N\'eel phase boundary. The transition from a parabolic form of the free energy to a double-well potential form signals the formation of the
local magnetic moment, which for both orbitals occurs approximately at the same value of the interaction ${U=3.55}$ (orb.\,1) ${U_2=3.65}$ (orb.\,2).
\label{fig:moment}}
\end{figure}

To distinguish between the Slater and Heisenberg mechanisms of the OSNT we perform calculations for the local magnetic moment along the lines of Ref.~\onlinecite{PhysRevB.105.155151}:
Excluding the contribution of the itinerant electrons, we study the local free energy ${\cal F}[M_1;M_2]$, which is associated with the behavior of the local magnetic moment $M_1$ and $M_2$ of each orbital (for details see Ref.~\onlinecite{PhysRevB.105.155151}).
Due to the decoupling of the magnetic fluctuations in the two orbitals, the free energy takes the form ${{\cal F}[M_1;M_2] = {\cal F}[M_1] + {\cal F}[M_2]}$.
The corresponding result for the free energy obtained along the N\'eel phase boundary is shown in Figure~\ref{fig:moment}.
Remarkably, despite of the decoupling, 
the local magnetic moment in both orbitals is formed almost at the same point at the N\'eel phase boundary (${U_1=3.55}$ for orb.\,1 and ${U_1=3.65}$ for orb.\,2) as depicted by arrows in Figure~\ref{fig:D-TRILEX}).
This effect cannot be explained by a single-band picture that does not account for the interorbital Coulomb interaction $U_{ll'll'}$.
This interaction does not couple the magnetic fluctuations of different orbitals but is responsible for spin-flip processes that couple different orbitals. 
These processes allow for Kondo screening of the local magnetic moment, which otherwise would have been formed at the narrow orbital at smaller values of $U$, by itinerant electronic fluctuations of the wide orbital.
As a result, in the intermediate coupling regime the OSNT is governed by the Slater mechanism, although the transition to the ordered N\'eel state first occurs for the wide orbital and not for the narrow orbital as in the weak coupling case.

To complete the story, we also perform AFM (two-sublattice) DMFT calculations for ${J=0}$. 
In this case, in the weak coupling regime, the N\'eel transition occurs first for the narrow orbital (${T=0.023}$, ${U=1.2}$, red ``$\boldsymbol{\times}$'' marker in Fig.~\ref{fig:D-TRILEX}) and at larger values of the interaction - for the wide orbital (${T=0.105}$, ${U=3.6}$, blue ``$\boldsymbol{+}$'' marker in Fig.~\ref{fig:D-TRILEX}).  
The two N\'eel temperatures coincide at ${T=0.029}$ and ${U=2.4}$.
Interestingly, we observe that at small values of the interaction DMFT underestimates the transition temperature, and the crossing point ${T_{N_1}=T_{N_2}}$ is shifted to smaller temperature and larger $U$ compared to the \mbox{D-TRILEX} result. 
At larger interaction $U$ DMFT strongly overestimates the transition temperature.
In addition, we perform both, DMFT and \mbox{D-TRILEX} calculations for a non-zero value of the Hund's coupling ${J=U/6}$.
As expected from Ref.~\onlinecite{PhysRevLett.129.096404}, in this case the OSNT transforms into an ordinary N\'eel transition that occurs simultaneously for both orbitals.
The corresponding N\'eel temperatures are shown in Fig.~\ref{fig:D-TRILEX} for ${U=1.2}$ and ${U=2.4}$ by black ``$\blacktriangle$'' (\mbox{D-TRILEX}) and black ``$\boldsymbol{\times}$'' (DMFT) markers.
We find that at both interaction strengths DMFT predicts higher N\'eel temperatures compared to \mbox{D-TRILEX}, which is consistent with single-orbital calculations~\cite{PhysRevX.11.011058, PhysRevLett.129.107202, PhD_Vandelli}.

\begin{figure}[t!]
\includegraphics[width=0.82\linewidth]{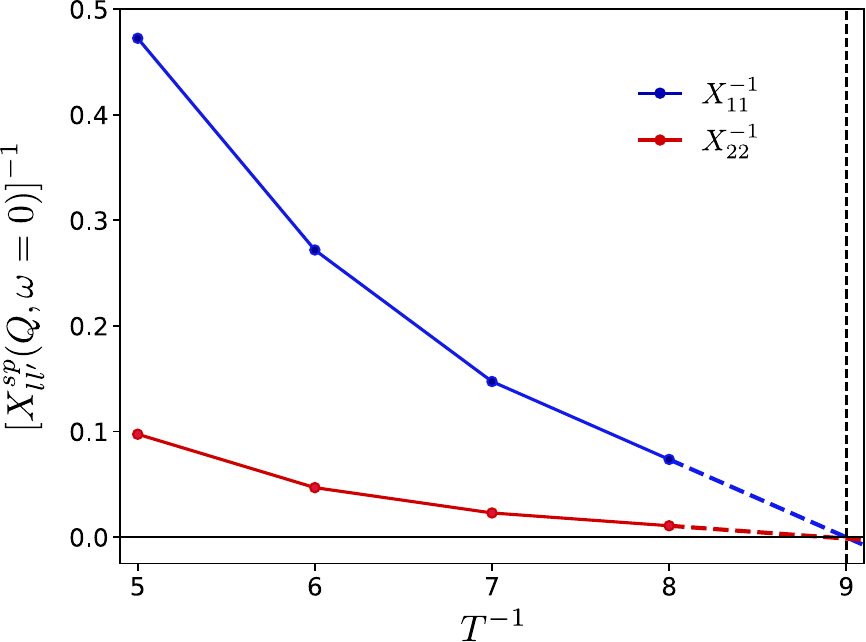}
\caption{Temperature dependence of the inverse of the $X^{sp}_{11}$ and $X^{sp}_{22}$ orbital components of the static (${\omega=0}$) spin susceptibility $X^{sp}_{ll'}$ calculated at the antiferromagnetic wave vector ${Q = \{\pi,\pi\}}$.
Results are obtained for the two-orbital Hubbard model on a square lattice considered in Ref.~\onlinecite{PhysRevB.85.035123} with ${t_1=t_2=1}$, ${t'_1=1}$, ${t'_2=0}$, ${U=4}$, and ${J=1}$.
The vertical dashed line at ${T^{-1}=9}$ indicates the N\'eel transition point at which both components of the spin susceptibility diverge.
The transition point was obtained by a linear extrapolation of the results (dashed blue and red lines).
\label{fig:X1234}}
\end{figure}

These findings motivate us to study the influence of long-range magnetic fluctuations on the magnetic \mbox{OSMIT} proposed previously in Ref.~\onlinecite{PhysRevB.85.035123} on the basis of DMFT calculations.
Here, we consider a two-orbital Hubbard model on a square lattice with nearest-neighbor ${t_1=t_2=1}$ and next-nearest-neighbor ${t'_1=1}$, ${t'_2=0}$ hopping amplitudes, Coulomb interaction ${U=4}$, and Hund's coupling ${J=1}$.
Within DMFT, for this set of model parameters the system lies well in the orbital-selective phase, characterised by localized N\'eel antiferromagnetic (AFM) behavior for the ${l=1}$ orbital, while the ${l=2}$ orbital remains itinerant~\cite{PhysRevB.85.035123}.
Within D-TRILEX, on the contrary, for this set of model parameters we do not observe any signature of an \mbox{OSMIT}, in line with the above findings results and Ref.~\onlinecite{PhysRevLett.129.096404}.
Instead, within \mbox{D-TRILEX} we observe a conventional (non-orbital-selective) N\'eel transition that occurs simultaneously for both orbitals.
Indeed, Figure~\ref{fig:X1234} shows that both orbital components of the AFM susceptibility diverge at the same critical temperature ${T_{N}\simeq0.1}$.
The critical temperature was obtained by a linear extrapolation (blue and red dashed lines in Fig.~\ref{fig:X1234}) of the inverse of each orbital component of the susceptibility as a function of the inverse temperature.
The absence of the orbital selective phase is thus due to the effect of long-range electronic correlations not captured in the cluster DMFT calculations of Refs.~\onlinecite{PhysRevB.81.220506, PhysRevB.84.020401, PhysRevB.85.035123, doi:10.1142/S0217984913300159}.

{\it Conclusions.}
In this work we have established strategies for realizing orbital-selective N\'eel-ordered magnetic states.
We have demonstrated that in the absence of Hund's exchange coupling $J$ the two-orbital Hubbard model with different bandwidths can indeed undergo an OSNT, at any interaction strength (except one specific value of the interaction).
Remarkably, the OSNT occurs differently in the weak and strong coupling regimes of interaction.
In the weak coupling regime it is governed by a Slater mechanism, namely in the absence of a local magnetic moment.
Consequently, the N\'eel transition occurs first for the narrow orbital, while the wide orbital remains itinerant.
In the strong coupling regime, a local magnetic moment is formed, and the Heisenberg mechanism of the OSNT leads to localized behavior occurring first in the wide orbital, while the narrow orbital stays itinerant.
Interestingly, at intermediate couplings we have found a non-trivial regime, where the transition occurs first for the wide orbital, but the local magnetic moment is not yet formed.
The latter is Kondo screened by electronic fluctuations of the wide orbitals. This results in a Slater mechanism of the OSNT at intermediate interaction strengths.
Most intriguingly, in the presence of Hund's coupling the OSNT is destroyed altogether: Hund's exchange effectively couples orbital degrees of freedom, and is thus detrimental
to orbital-selective behaviour. The ubiquity of Hund's exchange in real materials may provide a 
natural explanation for OSNTs likely being rather the exception than the rule. Nevertheless, a systematic
search for OSNP in real materials might prove worthwhile in view of potential applications in spintronics devices, 
e.g. for memory applications, spin valves or spin-charge converters.
Our work strongly suggests a materials screening among materials with as low as possible Hund’s exchange coupling.
A trivial corollary of this argument is obtained by replacing orbital indices by site indices:
thanks to the intrinsically weak direct intersite exchange interaction, 
site-selective magnetic orderings in materials
with several correlated sites per unit cell should be found more easily than orbital-selective OSNTs.

\begin{acknowledgments}
The authors acknowledge support from IDRIS/GENCI Orsay under project number A0130901393 and the help of the CPHT computer support team.
\end{acknowledgments}

\bibliography{Ref}

\end{document}